\documentclass[aps,prb,superscriptaddress,twocolumn,longbibliography]{revtex4-1}
\usepackage{amsmath}
\usepackage{amssymb}
\usepackage{graphicx}
\usepackage[usenames,dvipsnames]{xcolor}
\usepackage{tikz}
\usepackage{pgffor}
\usepackage{verbatim}
\usepackage{float}
\usepackage[colorlinks, breaklinks=true,linkcolor=red, citecolor=blue, linktocpage=true]{hyperref}\usepackage{cleveref}

\usepackage{color}

\def\nn{\nonumber}

\newcommand{\bea}{\begin{eqnarray}}
\newcommand{\eea}{\end{eqnarray}}
\newcommand{\la}{\label}

\newcommand{\be}{\begin{equation}}
\newcommand{\ee}{\end{equation}}



\begin{document}

\title{Dynamical many-body localization in an integrable model} 
\author{Aydin Cem Keser}
 \affiliation{Condensed Matter Theory Center, Department of Physics, University of Maryland, College Park, MD 20742, 	USA.}
 \author{Sriram Ganeshan}
 \affiliation{Condensed Matter Theory Center, Department of Physics, University of Maryland, College Park, MD 20742, USA.}
 \affiliation{Joint Quantum Institute, University of Maryland, College Park, MD 20742, USA.}
 \author{Gil Refael}
 \affiliation{Institute of Quantum Information and Matter, Department of Physics, California Institute of Technology, Pasadena, CA 91125, USA.}
\author{Victor Galitski}
\affiliation{Condensed Matter Theory Center, Department of Physics, University of Maryland, College Park, MD 20742, USA.}
 \affiliation{Joint Quantum Institute, University of Maryland, College Park, MD 20742, USA.}
 \affiliation{School of Physics, Monash University, Melbourne, Victoria 3800, Australia.}

\date{\today}

\begin{abstract}
We investigate dynamical many-body localization and delocalization in an integrable system of periodically-kicked, interacting linear rotors. The linear-in-momentum Hamiltonian makes the Floquet evolution operator analytically tractable for arbitrary interactions. One of the hallmarks of this model is that depending on certain parameters, it manifests both localization and delocalization in momentum space. 
We present a set of  ``emergent'' integrals of motion, which can serve as a fundamental diagnostic of dynamical localization in the interacting case. 
We also propose an experimental scheme, involving voltage-biased Josephson junctions, to realize such many-body kicked models.

\end{abstract}
\maketitle

\section {Introduction}
Recently, there has been a lot of interest and progress in understanding Anderson-type localization properties of disordered, interacting many-body systems. Notably, the remarkable phenomenon of many-body localization
(MBL) was discovered~\cite{Basko06,oganesyan2007,pal2010,moore2012, vadim, bauer2013area, bauer2014analyzing,imbrie2014many}. In an isolated system, MBL manifests itself in the localization of all eigenstates and leads to the breakdown of ergodicity and violation of the eigenvalue thermalization hypothesis~\cite{1991_Deutsch_ETH_PRA, 1994_Srednicki_ETH_PRE}, forcing to revisit the very foundations of quantum statistical mechanics~\cite{altman2014universal, huse2015many}.

In this work, we ask whether a driven interacting system can be {\em dynamically} many-body localized. We answer this question in the affirmative and present an exactly solvable model of a kicked chain of interacting linear rotors, which shows both dynamical MBL and delocalized regimes.  

A quantum kicked-rotor is a canonical model of quantum chaos~\cite{fishman, altland1996field} which exhibits dynamical localization in momentum space. The time dependent Schr\"odinger equation for a general kicked rotor is given by (here and below, we set $\hbar =1$ and the driving period $T=1$),
\begin{align}
\mathbf{i}\partial_t \psi(\theta,t)=[2\pi\alpha (-\mathbf{i}\partial_\theta)^l +K(\theta)\delta(t-n)]\psi(\theta, t)	.
\la{qkr}
\end{align}
In the ground breaking paper~\cite{fishman},  Fishman, Prange, and Grempel proved that the eigenvalue problem for the Floquet operator of a kicked rotor is equivalent to that of a particle hopping in a (quasi)periodic potential ((ir)rational $\alpha$) in one-dimension~\cite{prange} given by,
\begin{align}
\sum_r W_{m+r} u_r+\tan[\omega-2\pi\alpha m^l] u_m=E u_m.
\la{MM}	
\end{align}
This mapping traced the origin of the dynamical localization in driven systems to Anderson localization in time-independent settings. 
While the quadratic rotor ($l=2$) is nonintegrable, the linear rotor model ($l=1$) in Eq.~\eqref{qkr} was exactly solved in Refs.~\cite{prange,fishman, grempel,fishman1984localization,berry1984incommensurability}. The corresponding integrable lattice version in Eq.~\eqref{MM} with $l=1$ is dubbed Maryland model (MM)~\cite{Simon, watson} (See section \ref{sec:lattice_model} for technical details of this mapping). For the linear rotor, both classical and quantum dynamics is integrable, and the dynamical localization (absence of chaos) is due to the existence of a complete set of integrals of motion~\cite{berry1984incommensurability}. However, the incommensurate MM is an Anderson insulator with no classical interpretation~\cite{prange} even though the linear rotor manifests classical integrability. Thus, the dynamical localization for the quantum version of both linear ($l=1$) and quadratic rotor ($l=2$) seems to stem from the Anderson mechanism~\cite{prange}. 

 In this work, we generalize the linear kicked rotor model by considering an interacting chain of driven rotors in order to understand the anatomy of many-body localization in the dynamical space. One of the remarkable features of this many-particle model is that it manifests both localization and delocalization in dynamical space depending on whether the components of the $\vec \alpha$ are irrational or rational. 

 Let us emphasize from the outset that the model we consider is integrable for all parameters owing to the first order differential operator.  This leads to integrals of motion (IOMs), which are {\em local} in the spatial (angular) variables.  The underlying integrability is a special, non-universal feature of our model \eqref{ham}, which allows us to solve it exactly. However, the existence of the local-in-$\theta$ IOMs has no direct relation to dynamical MBL, which  occurs in {\em momentum} space. 
 
 As shown below, dynamical MBL is accompanied by the appearance of {\em additional} integrals of motion bounded in momentum space  -- a central result of this work. The non-interacting version of these additional IOM's for the linear kicked rotor was pointed out by Berry in Ref.[~\onlinecite{berry1984incommensurability}]. As argued below, these ``emergent'' IOMs and the dynamical MBL are  universal phenomena, which would survive in non-integrable generalizations of the model (which however is not analytically solvable).
 
To capture the dynamical localization for this interacting model we monitor three indicators: energy growth at long times, (momentum degrees of freedom at long times, and the existence of   integrals of motion in the momentum space. 
Below, we first present the detail of our model in Section~\ref{sec:model}. The main analytical results are outlined in Section~\ref{results}, and their numerical analysis and key conclusions are given in Section~\ref{sec:alphas}.  An outline of technical derivation of the results is given in Section~\ref{sec:derivations}.The experimental proposal to realize our model~\eqref{ham} is explained in Section~\ref{experiment}.
In section~\ref{conc} we list our conclusions and Section~\ref{ack} contains acknowledgements. Finally, Appendix~\ref{sec:lattice_model} and Appendix~\ref{sec:ddim_lattice} are devoted to the summary of the connection of the rotor problem to a lattice model and the correspondence between our model and a $d$-dimensional lattice, respectively.  
 
Throughout the text, in all the summations $\sum_{i}...$ or $\sum_{i j}...$ we only consider $i \ne j$. So that, expressions like $\sum_j 1/(\alpha_i -\alpha_j)$ are not divergent. The denominator vanishes only when there is a resonance, such as $\alpha_i \to \alpha_j $. 
 
\section {The Model}
\label{sec:model}

 We consider a many body interacting generalization,
\begin{align}
\hat{H}(t)&=\hat{H}_0+\hat{V} \sum^{\infty}_{n=-\infty}\delta(t-n) \nn,\, \mbox{  with } \hat{V}=\sum^{d}_{i=1} K(\hat{\theta}_i),\\
\hat{H}_0&=2\pi \sum\limits_{i=1}^d \alpha_i \hat{p}_i +\frac{1}{2}\sum^{d-1}_{i\ne j} J_{ij}(\hat{\theta}_i- \hat{\theta}_j)\label{ham}
\end{align}
of the linear rotor model.

$\hat{H}_0$ is the static Hamiltonian describing $d$ particles on a ring, each rotating $2\pi \alpha_i$ radians per one period of the kick. $\hat{\theta}_i$ is the position operator for the $i$th particle on the ring and $\hat{p}_i$ is its angular momentum operator, which has integer eigenvalues in the $\hbar=1$ units. These $d$ particles interact through a translationally invariant two-body potential $J_{ij}(\hat{\theta}_i - \hat{\theta}_j)$. This form of the interaction ensures conservation of momentum .  The rotors are driven by $\hat V(t)$ which contains periodic delta function impulses, where the strength of the impulse is given by a general periodic one-body potential $K(\hat{\theta}_i)$. The local potentials are periodic with $2\pi$ and therefore the most general form can be written as  $K(\theta_j) = \sum_m k_m e^{\mathbf{i}m\theta_j}$. Here, $k_m$ is the $m$th Fourier component of the potential that acts on the $j$th particle. The periodic form of the interaction potential $J$ can be written as $J_{ij}= \sum_m b^{ij}_m e^{\mathbf{i} m (\theta_i - \theta_j) }$. Here, $b^{ij}_m$ is the $m$th Fourier component of the interaction potential between the $i$th and $j$th particle. Reversing $i$ and $j$ and replacing $m$ with $-m$ in this Fourier expansion has no effect on the components, therefore $b_m ^{i j} = b_{-m}^{j i}$. This property will be handy while deriving formulas throughout the text. 
 
{Note that in our model the localization is  a consequence of incommensurate driving period and angular velocities $\vec \alpha$. This situation  is different from recent works using Floquet analysis to probe dynamical properties of  MBL states of disordered Hamiltonians~\cite{lazarides2014fate, ponte2014many, abanin2014theory, ponte2015periodically} (in these papers, a Floquet perturbation is imposed on a state, many-body localized in coordinate space, while our goal is to induce {\em dynamical} many-body localization in {\em momentum space} by the Floquet perturbation).}

\section {Main Results}
\label{results}

\subsection {Energy dynamics}
 Following the conjecture of D'Alessio and Polkovnikov~\cite{d2013many}, we test the dynamical localization by computing the energy  growth as a function of time at long times. The average energy after $N$ kicks (equivalent to time) can be written as,  $E(N)=\langle \psi_N|\hat{H}_0|\psi_N\rangle$. To compute it, we write
\begin{align}
|\psi_{N} \rangle = \hat{U}^N_F| \psi_0 \rangle, \ \hat{U}_F=e^{-\mathbf{i} \hat{V}} e^{-\mathbf{i} \hat{H}_0}, 
\end{align}
where $\hat{U}_F$ is the Floquet evolution operator, which captures the state of the system after each kick.  Owing to the linear dependence of the Hamiltonian on the momentum, we can explicitly compute this expectation. 
 \begin{align}
	&E(N)=E(0) +\sum^{d}_{i=1}\sum_m 2\pi \alpha_i\langle \hat{\Gamma}_{mi} \rangle_0\frac{\sin(m N\pi  \alpha_i)}{\sin(m\pi \alpha_i)}.
	\la{enav}
\end{align}
In the above expression, $E(0)=\langle \psi_0|H_0|\psi_0\rangle$ corresponds to the average many body energy over the initial state. $\hat{\Gamma}_{mi}=-\mathbf{i} m k_m e^{\mathbf{i} m \left( \hat{\theta}_i + \pi \alpha_i [N+1] \right)}$ depends on the chosen form of $K(\hat{\theta})$ averaged over the initial state and due to its periodic nature is a bounded function of the number of kicks $N$. The growth of energy for large $N$ is then completely dependent on the nature of $\alpha_i$ appearing in the ratio $\frac{\sin(m N\pi  \alpha_i)}{\sin(m\pi \alpha_i)}$. Note that this expression is completely independent of interactions, which we prove in Section~\ref{dav}. This is a consequence of momentum conserving interactions and this property is no longer valid when the translational invariance of interactions is broken.
\subsection {Momentum dynamics}
\label{momentum_dynamics}
 Another indicator for dynamical localization is the spread in the momentum degrees of freedom. 
    The $i$th momentum after $N$ kicks, $p_i(N)=\langle \psi_N|\hat{p}_i|\psi_N\rangle$ is given by the following expression,
\begin{align}
	p_i(N)=\langle \hat{p}_i \rangle_0+&\sum_m\langle \hat{\Gamma}_{mi} \rangle_0\frac{\sin(m N\pi  \alpha_i)}{\sin(m\pi \alpha_i)}\nn\\+&\sum_{m j }\langle \hat{\Gamma}^{int}_{mij} \rangle_0\frac{\sin(m N\pi  \Delta\alpha_{ij})}{m\pi\Delta\alpha_{ij}}.
	\la{pav}
\end{align}
We have defined $\Delta\alpha_{ij}=\alpha_i-\alpha_j$. In the above expression, the first term is the $i$th momentum in the initial eigenstate $\langle p_i \rangle_0=\langle \psi_0|\hat{p}_i|\psi_0\rangle$. The second term corresponds to the kicking potential as defined in Eq.~\eqref{enav}. The last term depends explicitly on the form of interaction via  $\hat{\Gamma}^{int}_{mij}=-\mathbf{i} m b_m^{ij} e^{\mathbf{i}m(\hat{\theta}_i-\hat{\theta}_j+\pi N \Delta\alpha_{i j})}$ and is a bounded function of $N$. The growth of momenta at long times corresponding to the last term is completely determined by the ratio  $\frac{\sin(m N\pi  \Delta\alpha_{ij})}{m\pi\Delta\alpha_{ij}}$.
\begin{figure*}[htp!] 
\textrm{ (AI) $\vec\alpha$ $\in$ irrational, $\alpha_i\ne \alpha_j$}\hspace{1cm}\textrm{(BI)  $\vec\alpha$ $\in$ irrational, $\alpha_1=\alpha_2$}\hspace{1.2cm}\textrm{(CI) $\alpha_1=1/2$, $\alpha_{2,..,10} \in$ irrational }\\
\includegraphics[width=5cm,height=4cm]{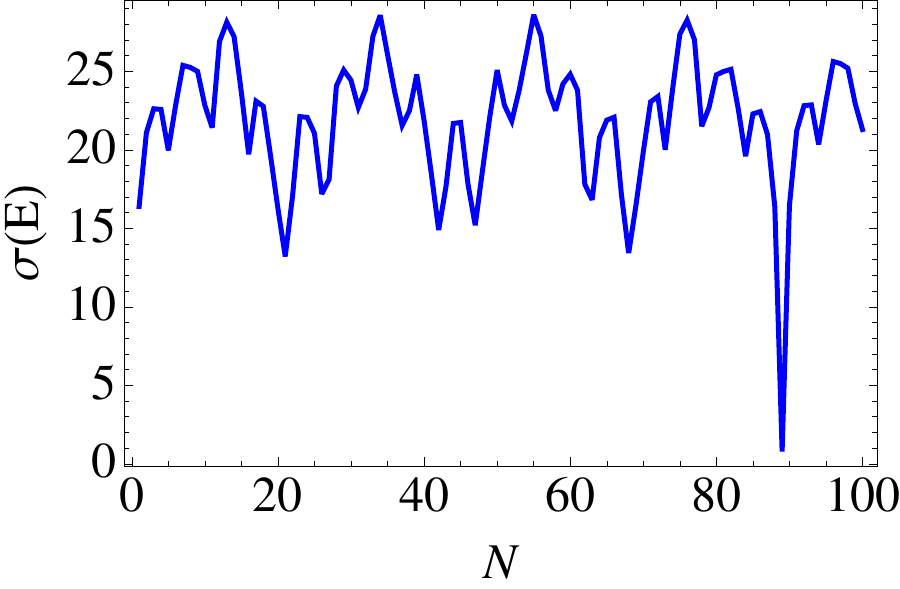}\hspace{0.1cm}\includegraphics[width=5cm,height=4cm]{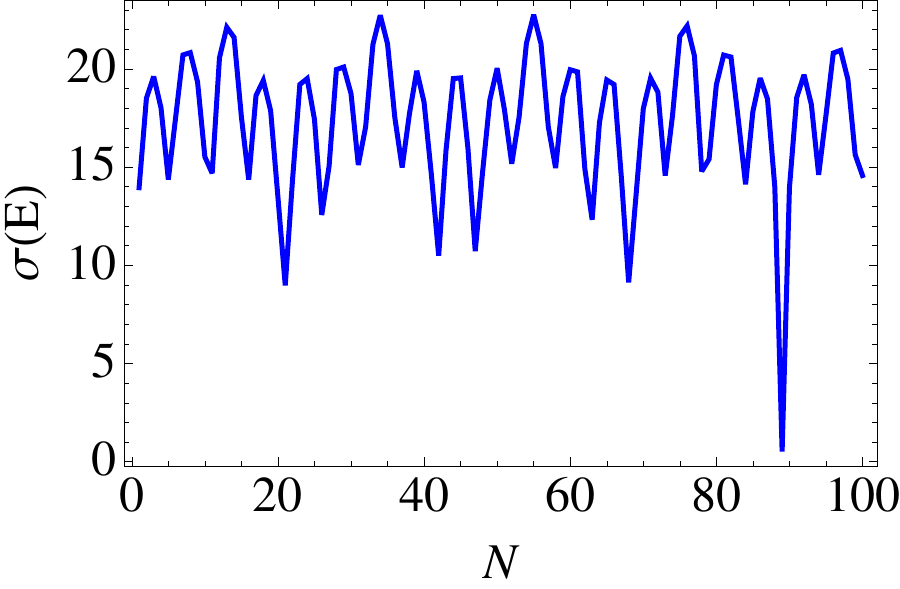}\hspace{0.1cm}\includegraphics[width=5cm,height=4cm]{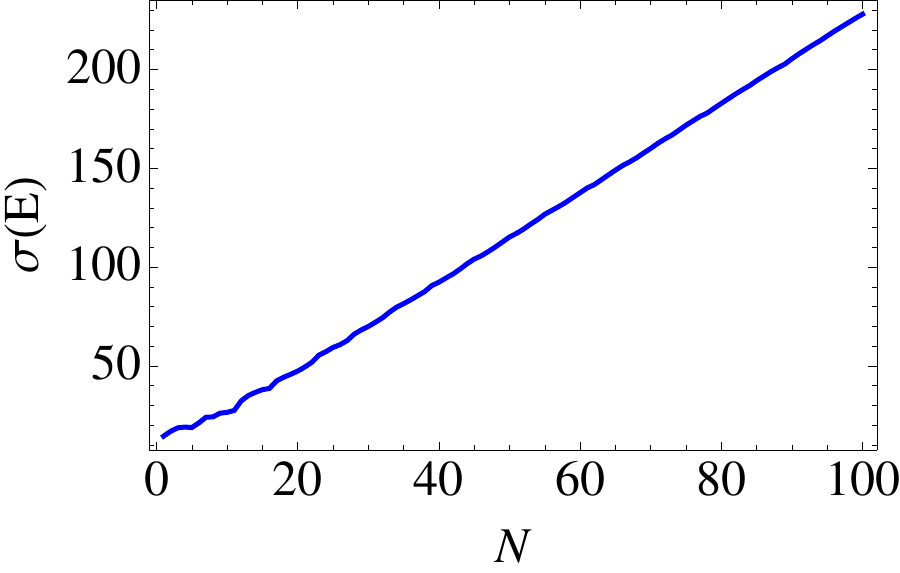}\\
\textrm{(AII) $\vec\alpha$ $\in$ irrational, $\alpha_i\ne \alpha_j$}\hspace{1cm}\textrm{(BII) $\vec\alpha$ $\in$ irrational, $\alpha_1=\alpha_2$}\hspace{1.2cm}\textrm{(CII) $\alpha_1=1/2$, $\alpha_{2,..,10} \in$ irrational }\\
\includegraphics[width=5cm,height=4cm]{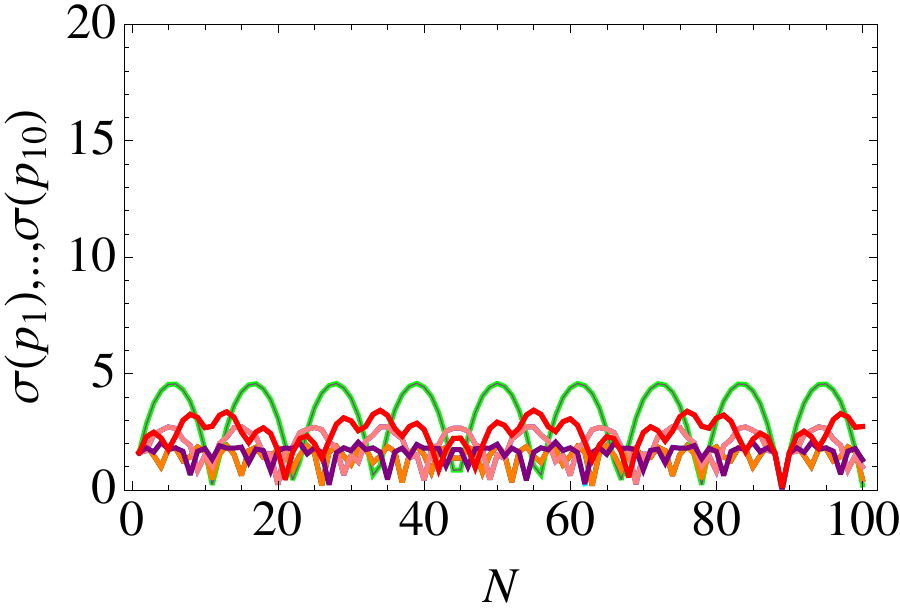}\hspace{0.1cm}\includegraphics[width=5cm,height=4cm]{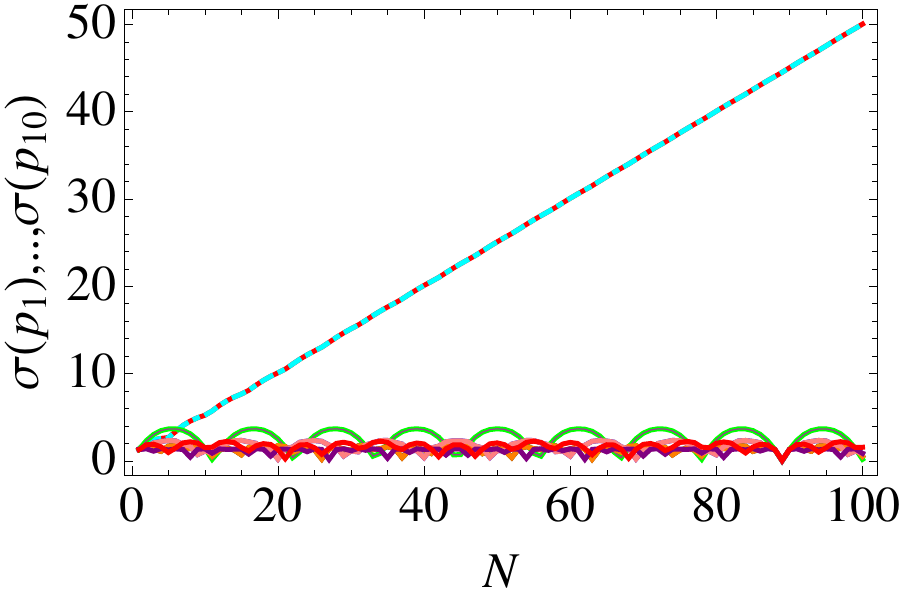}\hspace{0.1cm}\includegraphics[width=5cm,height=4cm]{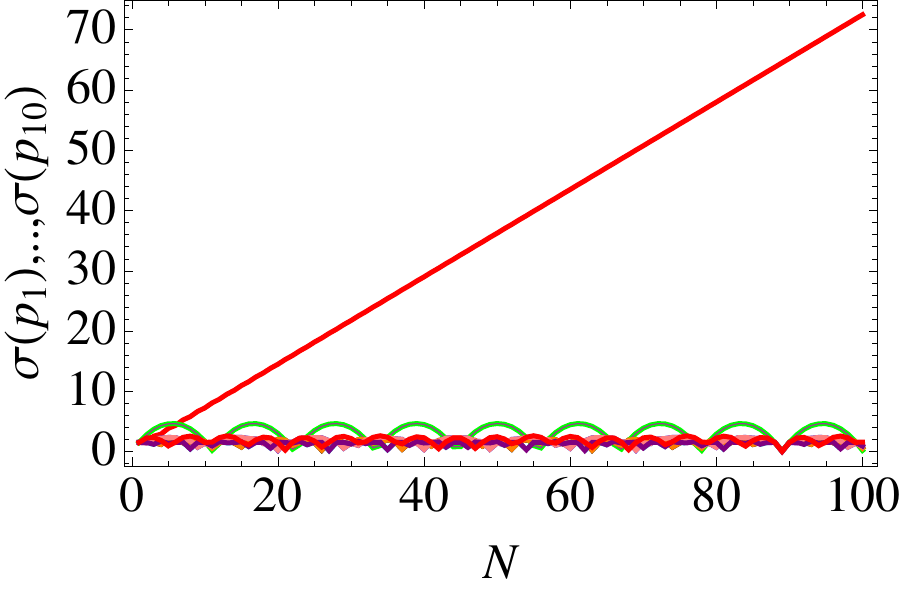}\\
\caption{(Top panel) Plots showing evolution of the root mean square deviation of energy, $\sigma(E)=\sqrt{\langle H^2_0 \rangle-\langle H_0 \rangle^2}$, as a function of time, labeled by the number of kicks, $N$. (Bottom Panel) Plots showing evolution of the root mean square deviation of individual momenta in an ensemble of 10 particles $\sigma(p_i)=\sqrt{\langle p^2_i \rangle-\langle p_i \rangle^2}$ as a function of time, labeled by the number of kicks, $N$. The initial states of the rotors are assumed to be definite momentum states. The total number of particles in this interacting ensemble is $d=10$ and circular boundary conditions apply. We consider two body interaction term to be $J_{ij}=\cos(\theta_i-\theta_j)$. The periodic kicking potential is given by $K(\theta)=\sum^{\infty}_{m=1}k_m \cos(m \theta)$ for all particles. The $m$-th Fourier coefficient is fixed by $k_m= \tilde{z}/m^2$, where $\tilde{z}$ is a random complex number with real and imaginary parts uniformly distributed in the interval [0,1]. We make sure that $K(\theta)$ is real. We choose $\vec \alpha$ corresponding to three scenarios:  Figs. (AI, AII):  If $\varphi$ denotes the golden ratio, $\alpha_j =   (j/d)\varphi-(1/2)\varphi$ are all irrationals for $j=1 \ldots 10$. Figs. (BI, BII): We consider $\alpha_j$  distinct irrationals as in Case A, but set $\alpha_1=\alpha_2$, which results in the resonant growth of $\sigma(p_1)$ and $\sigma(p_2)$, while $\sigma(E)$ is bounded.  Figs. (CI, CII): We consider $\alpha_j$ as in Case A  and set $\alpha_1=1/2$, which results in the resonant growth of $\sigma(E)$ and $\sigma(p_i)$.}
\label{fig:MBLFig} 
\end{figure*}

\subsection {Integrals of motion}
Recent works have shown that the existence of integrals of motion (IOM) can be used as a diagnostic to quantify both non-interacting Anderson localization~\cite{modak2015integrals, li2016quantum} and many body localization~\cite{serbyn2013, chandran2015, ros2015integrals}. In the context of dynamical localization for the model at hand, we work in the momentum basis and search for existence of IOMs in this basis. We begin by constructing IOMs by identifying operators $\hat{C}_i$ that satisfies $[\hat{C}_i,\hat{U}_F] = 0$ and $[\hat{C}_i, \hat{C}_j]=0$. The existence of these IOMs encode information about dynamical localization. Operators that commute with $\hat{U}_F$ satisfy the property $\langle \hat{C}_i \rangle_{N+1}=\langle \hat{C}_i \rangle_{N}$ and are given by
\begin{align}
\hat{C}_i =  \hat{p}_i + \frac{1}{2}\sum_m \frac{m k_m e^{\mathbf{i} m (\hat{\theta}_i + \pi \alpha_i)}}{\sin(m \pi \alpha_i)}+\frac{1}{2} \sum_{mj} \frac{b_m^{ij}e^{\mathbf{i} m (\hat{\theta}_{i}-\hat{\theta}_j )}}{\pi(\alpha_i-\alpha_j)}.
\la{lioms}
\end{align}
This expression is a generalization of the constant of motion given by Berry~\cite{berry1984incommensurability}.  The
 derivation of this expression is given in Sec.~\ref{subsec:consturct_IOM}. Since $\hat C_i$ is an IOM, $\hat p_i$ is bounded in time as long as  the series in the last two terms converges. The delocalization of $\hat p_i$ with time will occur due to the diverging denominators in the $\hat \theta$ dependent terms.  
 
We reiterate that the model always contains $d$ integrals of motion, $\hat B_i= \hat\theta_i/\alpha_i (\text{mod}\ 2\pi)$ ($[\hat B_i, \hat{U}_F]=0$) which results in integrability for both localized and delocalized cases. For example, if $\alpha_i = r_i/s_i$ is rational, the existence of $\hat B_i$ results in integrability even though $\hat C_{1 \ldots d}$ do not exist. Since $\hat B_i$ is momentum independent, its existence cannot bind $p_i(N)$.  For the fully localized case, we have additional $d$ IOM's $\hat C_{1 \ldots d}$ given in Eq.~\eqref{lioms} that explicitly depend on $\hat p_i$ and thus constrain it. Thus, $\hat C_{i}$'s can be thought of as ``emergent'' IOM's constraining the momentum growth, resulting in dynamical MBL and ``emergent'' superintegrability.

\section {Dynamical many-body localization and the structure of  $\vec \alpha$}
\label{sec:alphas}
Equipped with three analytical expression for the energy growth, momentum growth and IOMs [~\cref{enav,pav,lioms}],  we now diagnose the dynamical localization for three distinct cases with varying structure of $\vec \alpha$. We note that there are cases where $m \alpha_i$ can get arbitrarily close to integers (Liouville numbers) and total energy and momentum are no longer bounded.  In the single rotor case, such a situation yields interesting consequences like marginal resonance and a mobility edge in the momentum lattice, \cite{berry1984incommensurability, mobility_edge}. In this study we exclude this possibility and refer to generic irrationals only. 

\subsection {Case I: $\alpha_1 \ldots \alpha_d$ are distinct generic irrationals} 
\label{case1}
For irrational values of $\alpha_i$, the total energy in Eq.~\eqref{enav} is always a bounded function of $N$ since $m \alpha_i \notin \text{integer}$.  In Fig.~\ref{fig:MBLFig}(AI), we fix initial states to be momentum eigenstates $\psi_0(\vec{\theta}) $ which is $e^{\mathbf{i}\vec{ p_0}\cdot\vec{ \theta}}$ up to a normalization factor. For momentum eigenstates, $\langle\hat{\Gamma}_{mi}\rangle_0 = 0$ as the expectation value involves the integral $\int \text{d} \theta e^{\mathbf{i} m \theta}  $ over a circle. Thus, we plot the root mean square (RMS) of system's energy $\sigma(E)=\sqrt{\langle \hat{H}^2_0 \rangle-\langle \hat{H}_0 \rangle^2}$ as a function of $N$. Fig.~\ref{fig:MBLFig} (AI) shows boundedness in the spread of the energy as a function $N$. The momentum growth shown in Eq.~\eqref{pav} has contributions from both interactions and the kicking potential. In Fig.~\ref{fig:MBLFig} (AII) we plot RMS deviation of momenta $\sigma(p_1) \ldots \sigma(p_{d})$ (we used $d=10$ for the specific simulation) and show that  the spread in the momenta is bounded as a function of $N$. For this case the integrals of motion in Eq.~\eqref{lioms} exist and convergent. Thus, all the diagnostics for this case point towards a true many body dynamical localization. 
\subsection {Case II: $ \alpha_1 = \alpha_2 $ and $\alpha_2 \ldots \alpha_d$ are distinct generic irrationals} 
\label{subsec:case2} 
For this case, the energy remains a bounded function of $N$ since $m \alpha_i$ are not integers. Thus, the RMS deviation $\sigma(E)$ is bounded as shown in Fig.~(\ref{fig:MBLFig} BI). However, the second term in the momentum expression (Eq.~\eqref{pav}) develops a resonance (for the momenta $p_1$ and $p_2$) since $\alpha_1\rightarrow\alpha_2$. Due to this resonance term, $\sin(m N\pi  \Delta\alpha_{12})/(m\pi\Delta\alpha_{12}) \sim N$ as $\alpha_1\rightarrow\alpha_2$. This resonant growth is reflected in the RMS deviation of $\sigma(p_1)$ and $\sigma(p_2)$ growing linearly with $N$, while the $\sigma(p_3) \ldots \sigma(p_{10})$ remain bounded as shown in Fig.~\ref{fig:MBLFig} (BII). This is a striking scenario where the resonant momenta are not localized even if the total energy is bounded for large $N$. However, the delocalization of $p_1$ and $p_2$ in time is reflected in the break down of IOMs $\hat{C}_1$ and $\hat{C}_2$ due to diverging denominators  in the resonant limit $\alpha_1\rightarrow\alpha_2$. For this case, the bounded total energy fails to diagnose delocalization as shown in Fig.~\ref{fig:MBLFig} (BI, BII). We note that this scenario has no analogue in the non-interacting limit.  Notice that interactions we considered possess translational invariance, i.e. in the form $J(\theta_i -\theta_j)$ given in Eq.~\eqref{ham} and therefore interactions conserve momentum. The dichotomy between the energy growth and momentum growth is a result of conservation of momentum and the linear dependence of energy on momentum. If we allow interactions that break translational invariance, momentum is no longer conserved  and resonance due to interactions triggers unbounded growth (delocalization) of both momenta and the total energy.
\subsection { Case III: 
$\alpha_1=1/2$ and $\alpha_2 \ldots \alpha_d$ are distinct generic irrationals}  For rational $\alpha_1=1/2$, the system develops a different kind of resonance compared to Case II. We consider a kicking potential with  ($k_2 \ne 0$). The resonance condition $m \alpha_1 \in \text{integer}$ can be satisfied for $m=2$ and the ratio $\frac{\sin(m N\pi  \alpha_1)}{\sin(m\pi \alpha_1)}$ grows as $N$. It results in  energy growth and delocalization of momenta $p_1$ as shown in the RMS deviations in Fig.~\ref{fig:MBLFig} (CI, CII). This is a converse situation to Case II where the energy delocalizes with the delocalization of $p_1$ even though the momenta $p_2, p_3, \ldots p_{10}$ are bounded. This situation is again captured by the IOMs, where $\hat{C}_1$ does not exist, while $\hat{C}_2$ \ldots $\hat{C}_{10}$ are well defined and convergent as seen in Eq.~\eqref{lioms}. 

We can  generalize the above representative cases. For each rational $\alpha_i$, its integral of motion $\hat{C}_i$ breaks down and the corresponding momentum and energy diverge with time. For each pair $\alpha_i = \alpha_j$  both $\hat{C}_i$ and $\hat{C}_j$ break down and the corresponding momenta diverge in opposite directions, therefore $\hat{C}_i + \hat{C}_j$ is still an IOM  and the total momentum and energy of the pair are bounded.


Other than the behaviour of energy in Case II in Section~\ref{subsec:case2}, the rest of our analysis apply equally to the interactions that break translational symmetry. 
 

\section {derivation of main results}
\label{sec:derivations}
 Having established the physical understanding of localization for this model, we now sketch the brief derivation leading to the final results in \cref{enav,pav,lioms}. In the following we consider the expectation of a generic operator as a function of time,
 $X(N)=	\langle \psi_N|\hat{X}|\psi_N\rangle$. We write the explicit evolution of the many body wave function between two successive kicks, $|\psi_{N}\rangle=	e^{-\mathbf{i} \hat{V}}e^{ -\mathbf{i} \hat{H}_0}|\psi_{N-1}\rangle$. Notice that $\hat{H}_0=2\pi \vec{\alpha}\cdot \hat{\vec{p}}+\frac{1}{2}\sum^{d-1}_{i\ne j} J_{ij}(\hat{\theta}_i- \hat{\theta}_j)$ contains the many-body interaction term which may seem daunting, however, the linear momentum term allows a factorization in the Floquet operator.
The Baker-Campbell-Hausdorff formula  $\hat{Z} = \ln(e^{\hat{X}} e^{\hat{Y}})$ becomes tractable when  $[\hat{X},\hat{Y}] = s \hat{Y}$. In this case, the result simply reads $\hat{Z} =  \hat{X} + s \hat{Y}/ (1- e^{-s})$.
 Now fix $m$ and let $\hat{X} =\mathbf{i} 2 \pi (\alpha_1 \hat{p}_1 + \alpha_2 \hat{p}_2)$ and $\hat{Y} = \mathbf{i} \tilde{b}_m^{1 2} \exp(\mathbf{i}m[\hat{\theta}_1-\hat{\theta}_2])$ where $\Delta\alpha_{1 2} = \alpha_1 -\alpha_2$. The result of the commutator reads $[\hat{X},\hat{Y}] = \mathbf{i} 2 \pi \Delta \alpha_{1 2} m \hat{Y}$, precisely the tractable case discussed above. Replacing $s$ with $\mathbf{i} 2 \pi \Delta \alpha_{1 2} m$ in the formula $\hat{Z} =  \hat{X} + s \hat{Y}/ (1- e^{-s})$ and inverting both sides of the equality, we obtain the following factorization
\begin {multline}
\label{beginfactor}
e^{- \mathbf{i} 2 \pi \vec {\alpha}\cdot \vec{p} - \mathbf{i} \tilde{b}_m^{1 2} \exp(\mathbf{i} m [\theta_1 -\theta_2])}\\ = e^{- \mathbf{i} {b}_m^{1 2} \exp(\mathbf{i} m [\theta_1 -\theta_2])} e^{- \mathbf{i} 2 \pi \vec {\alpha}\cdot \vec{p}}.
\end {multline}

For this to hold, the Fourier coefficients must satisfy 
\begin{align}
\tilde{b}_m^{ij} = \frac{\sin(\pi m \Delta\alpha_{ij})} {\pi m \Delta\alpha_{i j}}b_m^{i j}e^{-\mathbf{i}m \pi \Delta\alpha_{ij}}.
\label{tilda_b}
\end{align}
This argument can be generalized to more particles and Fourier coefficients. Due to linearity summations over particle indices $i ,j$ and Fourier indices $m$ are introduced. All in all, we can factorize the evolution operator for our model in Eq.~\eqref{ham} as,
\begin{align}
|\psi_{N} \rangle = e^{-\mathbf{i} \hat{V} -  \frac{\mathbf{i}}{2}\sum_{ij}\tilde{J}_{ij}(\hat{\theta}_i- \hat{\theta}_j)}e^{-\mathbf{i} 2\pi \vec{\alpha}\cdot \hat{\vec{p}}} | \psi_{N-1} \rangle,
\la{singlekick1}
\end{align}
where we have defined the modified interaction term as,
\begin {subequations}
\begin {align}
	\tilde{J}_{ij}&=\sum_{m} \tilde{b}_m^{ij}e^{\mathbf{i}m(\theta_i-\theta_j)}\\
&=\sum_{m} \frac{\sin(\pi m \Delta\alpha_{ij})} {\pi m (\Delta\alpha_{ij})}b_m^{ij}e^{\mathbf{i}m(\theta_i-\theta_j-\pi \Delta\alpha_{ij})}.
\end {align}
\label{conversion}
\end {subequations}

The advantage of this factorization is that the operator $e^{-\mathbf{i} 2\pi \vec{\alpha}\cdot \hat{\vec{p}}}$ is a translation operator in the position basis. We can rewrite Eq.~\eqref{singlekick1} in the position basis by acting with $\langle \theta |$ from the left. 
We define $\langle \theta|\psi_{N} \rangle=\psi_N(\theta)$ and  express $\langle \theta| e^{-\mathbf{i} 2\pi \vec{\alpha}\cdot \vec{p}} | \psi_{N-1} \rangle=\psi_{N-1}(\vec{\theta}-2\pi \vec{\alpha})$. For a single kick we then have,
\begin{align}
\psi_N(\vec\theta)=e^{-\mathbf{i} V(\vec{\theta}) - \mathbf{i} \frac{1}{2}\sum_{ij}\tilde{J}_{ij}(\theta_i-\theta_j)}\psi_{N-1}(\vec\theta-2\pi\vec\alpha).
\end{align}
The above equation can be recursively iterated to yield,
\begin {multline}
\psi_{N}(\vec{\theta}) =e^{-\mathbf{i} \sum\limits_{ n=0}^{N-1}\left[ V(\vec{\theta}-2\pi n \vec{\alpha}) +\frac{1}{2}\sum_{i j} 
\tilde{J}_{i j}(\vec{\theta} -2\pi n \vec{\alpha}) \right]}\\\times \psi_0(\vec{\theta}-2\pi N \vec{\alpha}).
\la{wfn}
\end {multline}
Here $\tilde{J}_{ij}(\vec{\theta} -2\pi n \vec{\alpha})$ is a short hand notation for $\tilde{J}_{ij}( \theta_{i} - \theta_j -2\pi n \Delta\alpha_{ij})$.
\subsection {Derivation of the evolution of energy and momentum averages}
\label{dav}
Now consider a generic operator $\hat{X}\equiv X(\hat{p}_1...\hat{p}_d; \hat{\theta}_1...\hat{\theta}_d)$. The expectation value of this operator after $N$ kicks is
\begin{align}
\label{generic}
 X(N)=\!\! \int d \vec\theta\  	\psi^*_{N}(\vec{\theta})X\left(\frac{\partial}{d\theta_1}...\frac{\partial}{d\theta_d}; \theta_1...\theta_d \right)	\psi_{N}(\vec{\theta}).
\end{align}

If we substitute $\hat{X}=\hat{p}_k$, we get
\begin {align}
\label {momentum}
p_l(N)& = \langle \hat{p}_l \rangle_0- \nonumber \\
&\sum_{n=1}^{N} \left\langle \partial_{l}  V(\vec{\theta}+ 2 \pi n \vec{\alpha}) +\partial_{l} \frac{1}{2} \sum_{i j } \tilde{J}_{i j} (\vec{\theta} + 2 \pi n \vec{\alpha}) \right \rangle_0.
\end {align}
Here, $\partial_l \equiv \partial / \partial \theta_l$ and $\langle ... \rangle_0 \equiv \int d \vec{\theta} ...|\psi_0|^2$. The contribution from the kicking potential is:

\begin {align}
&\left\langle -\partial_{l} \sum_n V(\vec{\theta}+ 2 \pi n \vec{\alpha})\right\rangle_0
= \sum_m \langle \hat{\Gamma}_{m l} \rangle_0 \frac{\sin(m N\pi  \alpha_l)}{\sin(m\pi \alpha_l)}.
\label {kicking}
\end {align}

The contribution from the interaction potential is

\begin {align}
\left\langle-\partial_l \sum_n \frac{1}{2}\sum_{i j} \tilde{J}_{i j} (\hat{\theta}_i -\hat{\theta}_j +2\pi n [\alpha_i -\alpha_j])\right\rangle_0\\
=\sum_{j m} \langle \hat{\Gamma}^{int}_{mlj}\rangle_0 \frac{\sin(m N \pi \Delta \alpha_{l i})}{\pi m \Delta \alpha_{l i}}. \nonumber
\end {align}
Putting together above expressions, we obtain the expression for the momentum dynamics presented in Sec.~(\ref{momentum_dynamics}). 
\begin{align}
	p_i(N)=\langle \hat{p}_i \rangle_0+&\sum_m\langle \hat{\Gamma}_{mi} \rangle_0\frac{\sin(m N\pi  \alpha_i)}{\sin(m\pi \alpha_i)}+\nn\\&\sum_{m j i}\langle \hat{\Gamma}^{int}_{mij} \rangle_0\frac{\sin(m N\pi  \Delta\alpha_{ij})}{m\pi\Delta\alpha_{ij}}.
	\la{pav_explained}
\end{align}


Now we derive the expression for the energy growth. Substituting $\hat{X}=\hat{H}_0$ in Eq.~\eqref{generic}, we have
\begin {align}
E(N) = E(0) + \sum_{m i} 2 \pi \alpha_i \langle \hat{\Gamma}_{m i}\rangle_0  \frac{\sin(m N \pi  \alpha_l)}{\sin(m\pi \alpha_l)}\nonumber \\+ \frac{1}{2} \sum_{i j} \langle J_{i j} (\hat{\theta}_i-\hat{\theta}_j + 2\pi N \Delta \alpha_{i j} ) - J_{i j}(\hat{\theta}_i -\hat{\theta}_j) \rangle_0 \nonumber \\+ \sum_{i j m} 2 \pi \alpha_i \langle \hat{\Gamma}_{m i j}^{int}\rangle_0 \frac{\sin(m N \pi \Delta \alpha_{i j})}{m\pi \Delta \alpha_{i j}}.
\label{energy}
\end {align}
In the above equation, a cancellation occurs between the interaction terms in the last two lines of Eq.~\eqref{energy} owing to the 
following relation,
\begin {align}
\frac{1}{2} \sum_{i j} \langle J_{i j} (\hat{\theta}_i-\hat{\theta}_j + 2\pi N \Delta \alpha_{i j} ) - J_{i j}(\theta_i -\theta_j) \rangle_0\nonumber\\
=-\sum_{i j m} 2 \pi \alpha_i \langle \hat{\Gamma}_{m i j}^{int}\rangle_0 \frac{\sin(m N \pi \Delta \alpha_{i j})}{m\pi \Delta \alpha_{i j}}.
\end {align}

This completes the derivation of the energy growth shown in Sec.~(\ref{energy})
\begin{align}
	&E(N)=E(0) +\sum^{d}_{i=1}\sum_m 2\pi \alpha_i\langle \hat{\Gamma}_{mi} \rangle_0\frac{\sin(m N\pi  \alpha_i)}{\sin(m\pi \alpha_i)}.
	\la{enav_explained}
\end{align}

The dropping out of interaction terms from the total energy can be interpreted in the following way. As seen from Eq.~\eqref{pav_explained}, the contribution of interaction to the momentum average picks up a negative sign when $i$ and $j$ are interchanged.  This means interaction transfers momentum from one particle to the other at each kick, in other words momentum is conserved for each couple of rotors. Since the energy is linear in momenta, when the momenta of rotors are summed, the contribution of interactions vanishes. We emphasize that when the interactions break translational invariance, they no longer conserve momentum and in that case energy growth depends on interactions too.

\subsection {Construction of integrals of motion}
\label{subsec:consturct_IOM}

In this section, we outline the derivation involved in the construction of integrals of motion. By inspecting Eq.~\eqref{momentum} and using the identity $\sin(m\pi \alpha) = [\exp(\mathbf{i} m\pi \alpha) - \exp(-\mathbf{i} m \pi \alpha)]/(2 \mathbf{i})$, we can write
\begin {multline}
p_l(N+1) - p_l(N) =\\   \frac{1}{2}\sum_m  \frac{m k_m}{\sin(m \pi \alpha_l)} \left( \left\langle e^{\mathbf{i}m (\hat{\theta}_l  +\pi \alpha_l)}\right\rangle_N -\left\langle e^{\mathbf{i}m (\hat{\theta}_l
+\pi \alpha_l)} \right \rangle_{N+1}\right)\\
+\frac{1}{2}\sum_{m j}  \frac{ b_m^{l j}}{\pi \Delta\alpha_{l j}}  \left(  \left\langle e^{\mathbf{i}m (\hat{\theta}_l-\hat{\theta}_j )}\right\rangle_N -
\left\langle e^{\mathbf{i}m (\hat{\theta}_l -\hat{\theta}_j)
} \right \rangle_{N+1}\right),
\end {multline}
noting that, the expression is valid whenever the denominators $\sin(m\pi\alpha_l)$ and $\Delta \alpha_{l j}=\alpha_l -\alpha_j$ are non-zero, in other words, whenever the resonances are avoided. The above expression can be organized in a way that it manifests the IOMs,
\begin{align}
	\left\langle\hat{p}_i + \frac{1}{2}\sum_m \frac{m k_m e^{\mathbf{i} m (\hat{\theta}_i + \pi \alpha_i)}}{\sin(m \pi \alpha_i)}+\frac{1}{2} \sum_{mj} \frac{b_m^{ij}e^{\mathbf{i} m (\hat{\theta}_{i}-\hat{\theta}_j )}}{\pi(\alpha_i-\alpha_j)}\right \rangle_{N+1}=\nonumber\\
	\left\langle\hat{p}_i + \frac{1}{2}\sum_m \frac{m k_m e^{\mathbf{i} m (\hat{\theta}_i + \pi \alpha_i)}}{\sin(m \pi \alpha_i)}+\frac{1}{2} \sum_{mj} \frac{b_m^{ij}e^{\mathbf{i} m (\hat{\theta}_{i}-\hat{\theta}_j )}}{\pi(\alpha_i-\alpha_j)}\right \rangle_{N}	
\end{align}

If we use the definition in Eq.\eqref{lioms} we get $\langle \hat{C}_l \rangle_{N+1} = \langle \hat{C}_l \rangle_{N} $, thereby proving that $\hat{C}_l$ is an integral of motion.

\subsection {Floquet Hamiltonian and quasienergy eigenstates}
\label{sec:eigenstates}

A special case of our model is when there are no resonances, namely, all IOMs associated with the momentum localization $\hat{C}_{1..d}$ are intact.  In this case, of particular importance is the following combination of the integrals of motion
\begin {equation}
\label{ham_expand}
\hat{H}_F = \sum_i 2\pi \alpha_i \hat{C}_i.
\end {equation}
Where $\hat{H}_F$ is known as the Floquet Hamiltonian and is defined as,
\begin {equation}
e^{-\mathbf{i}H_F} = e^{-\mathbf{i} \hat{V}} e^{-\mathbf{i} \hat{H}_0} = \hat{U}_F.
\end {equation}
The quasienergy  wavefunctions $\psi_\omega$ are simultaneous eigenstates of $\hat{H}_F$ and $\hat{C}_i$'s. The quasienergy-$\omega$ state centered around  momenta $\langle\hat{\vec{p}}\rangle=\vec{M}$ is

\begin {multline}
\label{eigen}
\psi_{\omega}(\vec{\theta}) =  (2\pi)^{-{N}/{2}} 
\exp \bigg\lbrace \mathbf{i} \vec{M}\cdot \vec{\theta} 
-  \sum_{j m} \frac{k_m e^{\mathbf{i} m (\theta_j + 
\pi \alpha_j)}}{2  \sin(m \pi \alpha_j)}  \\ -\sum_{i j m} \frac{b_m^{ij}}{  4\pi   m (\alpha_i-\alpha_j)} e^{\mathbf{i} m (\theta_i-\theta_j)}\bigg\rbrace.
\end {multline}

This satisfies $\hat{C}_i \psi_{\omega} = M_i \psi_{\omega}$. By writing $\hat{H}_F \psi_{\omega} = \omega \psi_{\omega}$ and using Eq.~\eqref{ham_expand}, we see that the eigenvalue equation is satisfied when, $\omega = 2\pi\vec{\alpha}\cdot \vec{M} (\text{mod}\ 2 \pi)$.

We can also compute the momentum-momentum correlator, $\langle \hat{p}_i \hat{p}_j\rangle - \langle \hat{p}_i\rangle \langle \hat{p}_j\rangle$  for $i \neq j$ over quasienergy eigenstates. The correlator over this state follows as
\begin {equation}
\label{corr}
 \langle \hat{p}_i \hat{p}_j \rangle - \langle \hat{p}_i\rangle \langle \hat{p}_j \rangle = -\frac{1}{4} \sum_m \frac{|b_m^{i j}|^2} { \pi^2 (\alpha_i - \alpha_j)^2}.
\end {equation}
In the case of a resonance $\alpha_i \to \alpha_j$, this correlator clearly diverges, while in the localized case it remains finite. 

\section {Experimental proposal}
\label{experiment}
\begin{figure} 
\includegraphics[width=0.5\textwidth ]{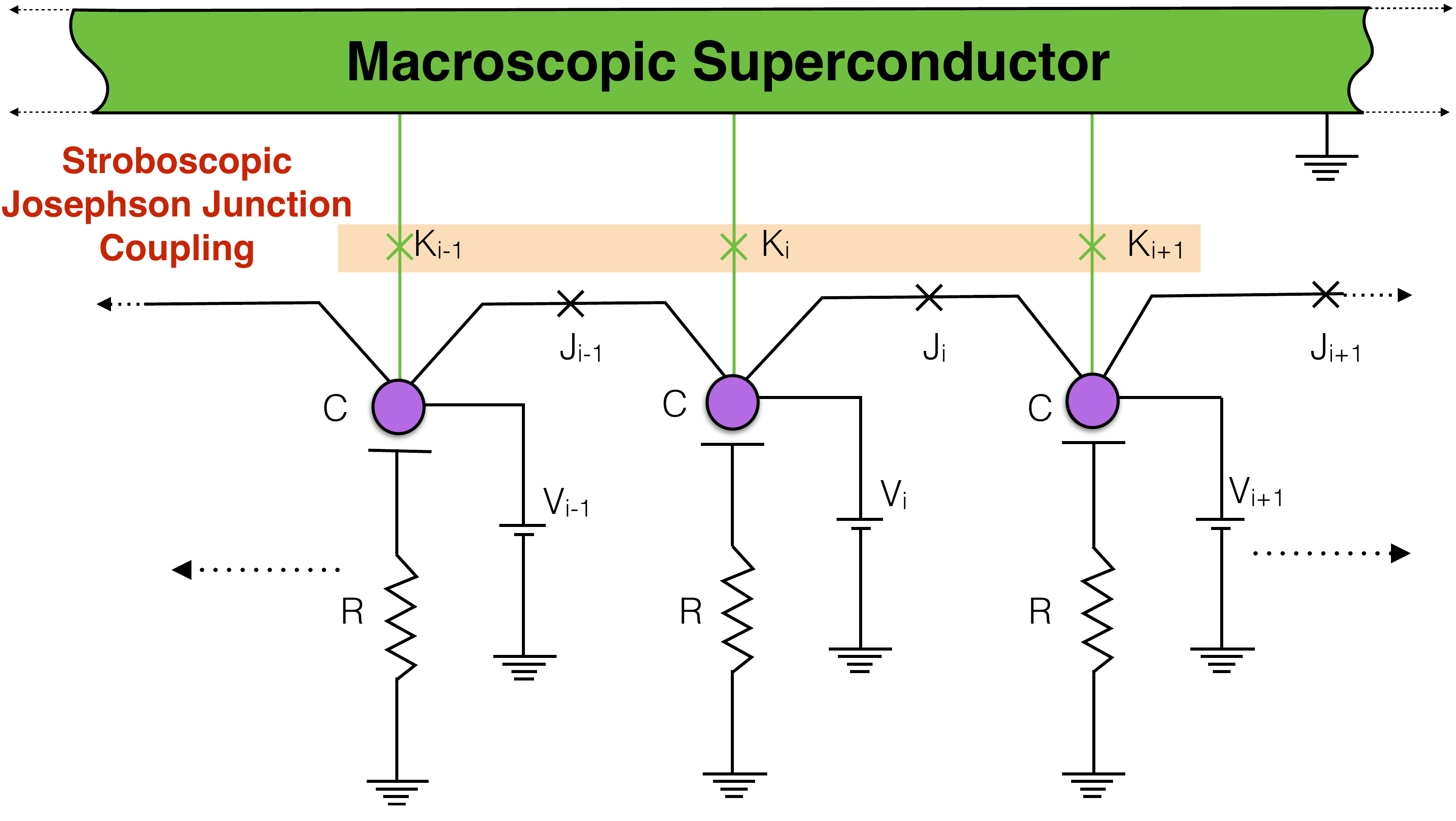}\\
\caption{Schematic of chain of voltage-biased ($V_i$) superconducting grains coupled to each other ($J_i$) using Josephson junctions. Each grain is connected ($K_i$) to a grounded common macroscopic superconductor stroboscopically through a strong Josephson coupling.}
\label{expfig} 
\end{figure}

 A natural venue for realizing the interacting kicked rotor model is superconducting grains. The Hamiltonian, Eq.~\eqref{ham} could be implemented using a chain of voltage-biased superconducting grains coupled to each other using Josephson junctions. Consider a chain of grains gated by a ground plane which is resistively connected to ground (see Fig.~\ref{expfig}). We then supply a gate voltage $V_i$ to each grain, and connect them to each other by Josephson junctions $J_{ij}$. 

 Because the voltage on the $i$-th grain  is locked to be $V_i$, its phase winds with an angular velocity $\dot\phi
 _i=2e V_i/\hbar$. Therefore, the resistance and gate capacitor can be ignored when writing the effective stationary part of the Hamiltonian:
\begin{equation}
H_0=  \sum_i q_i V_i -\sum_{ij}J_{ij}\cos(\phi_i-\phi_j).
\end{equation}
In addition, the ``kick'' term is obtained by connecting each grain to a common macroscopic superconductor (which is itself grounded), for a short time and through a strong Josephson coupling.
\be
V(t)=-\sum_{i,\,n} K(t-nT)\cos\phi_i,
\ee
with $K(t)=K$ when $|t|<\tau$. To make the kick term as close to a delta function as possible, we must have $ 2eV_i \tau\ll \hbar 2\pi$, and $K\tau\sim \hbar$, and  $\tau\ll T$. A diagram of the circuit for a nearest-neighbor interaction is shown in Fig.~\ref{expfig}. Identifying, $\phi_i=\theta_i$, $2 \pi \alpha_i=2e V_i /\hbar$ and $p_i=\hbar q_i/2e$ we see that we indeed obtain the Hamiltonian of Eq.~\eqref{ham}.

\section {Conclusions}
\label{conc}
In this work, we introduced the concept of {\em dynamical} many-body 
localization and presented an exactly-solvable model of driven 
linear rotors, which exhibits this phenomenon. Although the model 
possesses a full set of integrals of motion, it is shown that 
dynamical MBL is accompanied by the emergence of additional 
integrals of motion, local in momentum space. We believe that this observation  has important general implications for understanding dynamics of interacting many-body systems.

We have shown that these integrals of motion break down due to two types of resonances 
indicating delocalization in momentum space. One type of resonance 
originates from commensuration of the external driving period with 
the parameters of the system and the other from the static 
interactions. An interesting  feature of this model is that the 
total energy in the system, that is linear in momenta,  fails to be 
a good indicator of dynamical localization, since when 
momentum conserving interactions delocalize momentum, momenta of interacting pair grow in  opposite 
directions. 

Moreover, we have shown, (see Appendix) by utilizing the lattice mapping introduced by Fishman and Grempel and Prange, that our model maps into a disordered lattice with as many dimensions as there are rotors. Based on this observation, we argue that what is observed is an Anderson type localization, particularly of the type seen in correlated disorder systems. 

We also have proposed an experimental setup composed of Josephson junctions and superconducting grains to realize the model Hamiltonian. 

Finally, we emphasize that the results presented here can apply to more generic non-integrable systems. For example, a recent 
work~\cite{Rozenbaum} considers interacting kicked Dirac particles with individual Hamiltonians, $H_0 = 2 \pi \alpha \sigma_x p + M \sigma_z$, 
and provides a simple argument that this non- integrable system
also exhibits MBL. First, this model also exhibits localization when $\alpha$'s are generic distinct irrationals.  Second, although the number of interacting particles  that can be considered numerically is limited, note 
that at large momenta the Dirac model crosses over to the linear 
model considered here. This suggests that MBL should be robust to a class of non-integrable generalizations, for any number of interacting rotors. 

\section {Acknowledgements}
\label{ack}
 This work was supported by US-ARO, Australian Research Council, and Simons Foundation (A.C.K. and V.G.). SG gratefully acknowledges support by NSF-JQI-PFC and LPS-MPO-CMTC. GR is grateful for support from the IQIM and the Moore Foundation, and the NSF under grant DMR-1410435. The authors are grateful to Sankar Das Sarma and Efim Rozenbaum for useful discussions. 
 
\appendix
\section {Mapping between Quantum Kicked Rotor and the tight-binding model}

 In the first part of this section, we derive the known lattice mapping of the one dimensional kicked rotor~\cite{prange}. Once we establish this derivation, we show that the many-body linear kicked rotor \eqref{ham} also admits a lattice mapping of a particle on a d-dimensional lattice. We emphasize that existence of such a mapping in the many-body case is limited to the case of linear $p$ model ($l=1$). 

\subsection{Lattice model of single quantum kicked rotor}
\label{sec:lattice_model}
In the introduction we considered the time dependent Schr\"odinger equation for a kicked rotor. The kinetic part was considered to be $\hat{p}^l$. For the quadratic kicked rotor case $l=2$ and the linear kicked rotor that we consider in this work is $l=1$. Also $\hbar = 1$ and kicking period $T=1$ are used in this equation.
\begin{align}
\mathbf{i}\partial_t \psi(\theta,t)=[2\pi\alpha (-\mathbf{i}\partial_\theta)^l +K(\theta)\sum_n \delta(t-n)]\psi(\theta, t)	.
\la{qkr2}
\end{align}
The above equation can be solved for $\psi(\theta,t)=e^{-i \omega t}u(\theta,t)$, where the function $u$ has the unit periodicity of the driving. Let $u^{\pm}$ defines the state just before and after the kick and are connected by the evolution operator in the following way,
\begin{align}
u^+=e^{-\mathbf{i} K(\theta)}u^-,\quad u^-=e^{\mathbf{i} (\omega-2\pi \alpha \hat p^l)}u^+	
\end{align}
Define the following: $\bar u=\frac{u^++u^-}{2}$,  $\exp(-\mathbf{i} \hat{K}(\theta)) = (1-\mathbf{i} \hat{W}(\theta))^{-1}(1+\mathbf{i} \hat{W}(\theta))$, $\exp(-\mathbf{i}[ 2\pi \alpha \hat{p}^l-\omega]) = (1-\mathbf{i}\hat{T}(\theta))^{-1}(1+\mathbf{i} \hat{T}(\theta))$. Based on the above definitions,  $ u^{\pm}=(1\mp \mathbf{i}\hat T(\theta))\bar u$ and 
\begin{align}
[\hat T(\theta)+\hat W(\theta)]\bar u=0	
\end{align}
is obtained.

Fourier transforming the above expression, we get the following tight binding model,
\begin{align}
\sum_{n\ne m} W_{m-n} u_n+T_m u_m=E u_m.
\la{MM_single}	
\end{align}
Here, the energies and hoppings are:
\begin {subequations}
\begin {align}
\begin {split}
W_{m -n}&= - E \delta_{m, n}-\\ &\frac{1}{2\pi} \int_0^{2\pi}  e^{-\mathbf{i} (m - n) \theta}\left\lbrace\tan\left(\frac{K(\theta)}{2}\right)\right\rbrace \text{d}\theta,
\end {split}\\
T_{m}&=\tan\left(\frac{1}{2}[\omega - 2\pi \alpha  m^l]\right).
\end {align}\
\label{r_to_l} 
\end {subequations}
This completes the derivation for the lattice mapping for the single rotor case. In the following section, we generalize this derivation to demonstrate the existence of a lattice mapping for the interacting rotor model of Eq.~\eqref{ham}.
\subsection {$d$-dimensional lattice model}
\label{sec:ddim_lattice}
In this section we show that there exists a $d-$ dimensional lattice model corresponding to the $d$ particle interacting version of the kicked rotor model in Eq.~\eqref{ham}. Such a mapping has been previously identified for the case of a $d$ rotors driven by an interaction potential~\cite{fishman1984localization}. Notice that in our case, the interactions are encoded in the stating Hamiltonian $\hat H_0$. However, we show that for the linear momentum dependence of the kinetic term, static interactions can expressed as the driven interactions and the rest of the lattice mapping simple follows from Ref.~[\onlinecite{fishman1984localization}]. In order to establish this mapping, we use the factorization of the Floquet operator discussed in Sec.~(\ref{sec:derivations}), see \crefrange{beginfactor}{wfn}. The time dependent Hamiltonian in Eq.~\eqref{ham} produces the same Floquet operator as, 
\begin {equation}
\hat{H}(t)=2\pi \sum\limits_{i=1}^d \alpha_i \hat{p}_i+\hat{V}^F \sum^{\infty}_{n=-\infty}\delta(t-n),
\end {equation}
where we have defined, 
\begin {align}
\label{expanded}
\hat{V}^F=\sum^{d}_{i=1} K(\hat{\theta}_i) +\frac{1}{2}\sum_{i\ne j} \tilde{J}_{ij}(\hat{\theta}_i- \hat{\theta}_j) \\= \sum^{d}_{j=1} \sum_m k_m e^{i m \hat{\theta}_j} + \frac{1}{2}\sum_{i\ne j} \sum_m \tilde{b}_m^{i j } e^{i m (\hat{\theta}_i - \hat{\theta}_j)}.
\end {align}
The factorization enables to treat on site and interaction potentials on equal footing. The $\tilde{b}^{ij}$ is defined in Eq.~\eqref{tilda_b}.  Moreover, we write:
\begin {equation}
\label{multid}
\hat{V}^F = \sum_{\vec{m}} V^F_{\vec{m}} e^{i \vec{m} \cdot {\vec{\theta}}}.
\end {equation}

The equivalence between Eq.~\eqref{expanded} and Eq.~\eqref{multid} is conceptually very simple, but somewhat harder to put into words. It is probably best to give an example. Take the two rotor case i.e. $d=2$. Now $\vec{m}$ are two dimensional vectors with integer components. Let us fix $m \in \mathbb{Z}$. If $\vec{m} =  (m, 0 )$ then $V^F_{\vec{m}} = k_m$.  If $\vec{m} =  (0, m )$ then $V^F_{\vec{m}} = k_m$. If $\vec{m} =  (m, -m )$ then $V^F_{\vec{m}} = b_m^{1 2}/2$. For all other vectors, $V_{\vec{m}}^F = 0$. The summation in Eq.~\eqref{multid} is over all possible vectors $\vec{m}$.  Generalizing this notation to $d$ dimensions, it is possible for both the differences of angles and angles themselves to appear as $\vec{m} \cdot\vec{\theta}$.  Treating $\alpha$'s and $p$'s as $d$-dimensional vectors as well, the Hamiltonian can be succinctly as:

\begin {equation}
\hat{H}(t) = 2 \pi \vec{\alpha}  \hat{\vec{p}} +  \sum_{\vec{m}} V_{\vec{m}} e^{i \vec{m} \cdot {\hat{\vec{\theta}}}} \sum^{\infty}_{n=-\infty}\delta(t-n).
\end {equation}
Following Ref.~[\onlinecite{fishman1984localization}], the above driven Hamiltonian can be mapped on to a $d$-dimensional lattice model, which is closely related to the lattice mapping outlined in the previous section for the single rotor case.  
\begin {equation}
H_{{\vec{m},\vec{n}}} u_{\vec{n}} = T_{\vec{m}} u_{\vec{m}} + \sum_{\vec{n}} W_{\vec{m},\vec{n}} u_{\vec{n} } = E u_{\vec{m}}.
\label{lattice}
\end {equation}
 Here, $\vec{m}$ and $\vec{n}$ are vectors that contain integers that correspond to the quantized eigenvalues of the angular momentum operator. The hopping and onsite terms are defined as,
\begin {subequations}
\begin {align}
\begin {split}
W_{\vec{m} -\vec{n}}&= - E \delta_{\vec{m}, \vec{n}} + \\ &\frac{1}{(2\pi)^d} \int_0^{2\pi}  e^{-\mathbf{i} (\vec{m} - \vec{n})\cdot \vec{\theta}}\left\lbrace-\tan\left(\frac{V^F(\vec{\theta})}{2}\right)\right\rbrace \text{d}\vec{\theta}
\end {split},\\
T_{\vec{m}}&=\tan\left(\frac{1}{2}[\omega - 2\pi \vec{\alpha} \cdot \vec{m}]\right).
\end {align}
\label{r_to_l2} 
\end {subequations}
Here, we defined $E= -(1/[2\pi]^d) \int \tan\left(V^F(\vec{\theta})/2\right)$ so as to make $W_0 = 0$. 
\bibliography{references}
\end{document}